\documentclass[11pt,twoside]{article}
\usepackage[cp1251]{inputenc}
\usepackage{amssymb,amsmath,amscd}
\usepackage{fancyheadings}
\usepackage[russian]{babel}
\usepackage{bm}
\usepackage{graphicx}
\makeatother
\usepackage{wrapfig}
\newcommand{\GOD}{2023}
\newcommand{\UDK}[1]{\noindent{\footnotesize\sl УДК #1}}
\newcommand{\Nazva}[1]{\begin{center}\baselineskip=6.0mm{\Large\textbf{#1}}\end{center}\vspace*{0.5mm}}
\newcommand{\Avtor}[1]{\centerline{\large\textbf{\copyright~\GOD~г. \ #1}}\vspace*{-4.0mm}}
\newcommand{\AVTOR}{~}
\newcommand{\NAZVA}{~}
\newcommand{\lit}[3]{\vspace*{0.7mm}\par\noindent\makebox[5.2mm][r]{#1.~}\parbox[t]{159.8mm}{{\textit{#2}}~{#3}}\hspace*{-1.6mm}}
\begin{document}

\renewcommand{\abstractname}{}

\centerline{\large\textbf{УРАВНЕНИЯ С ЧАСТНЫМИ ПРОИЗВОДНЫМИ}}

%
%
%

\vspace{2mm}
\hrule
\vspace{2mm}

\UDK{517.956}
\Nazva{ТИПИЧНЫЕ ПРОВАЛЬНЫЕ АСИМПТОТИКИ КВАЗИКЛАССИЧЕСКИХ ПРИБЛИЖЕНИЙ К РЕШЕНИЯМ НЕЛИНЕЙНОГО УРАВНЕНИЯ ШРЁДИНГЕРА}
\Avtor{C. H. Мелихов, Б. И. Сулейманов, А. М. Шавлуков}
\renewcommand{\NAZVA}{ПРОВАЛЬНЫЕ АСИМПТОТИКИ КВАЗИКЛАССИЧЕСКИХ ПРИБЛИЖЕНИЙ}
\renewcommand{\AVTOR}{Мелихов, Сулейманов, Шавлуков}

\thispagestyle{empty}

\begin{abstract}\noindent
Обоснованы формальные асимптотики, описывающие типичные провальные особенности сборки квазиклассических приближений к решениям двух вариантов интегрируемого  нелинейного уравнения Шрёдингера $-i\varepsilon\Psi'_{t}=\varepsilon^2\Psi''_{xx}\pm2|\Psi|^2\Psi$,  где $\varepsilon$ ---  малый параметр. При обосновании используются идеология и факты математической теории катастроф, а также часть теоремы Ю. Ф. Коробейника, касающаяся аналитических при $h\to 0$ решениях $G(h,u)$ линейного уравнения смешанного типа $hG''_{hh}=G''_{uu},$ которому эквивалентны образы годографа обоих вариантов  систем уравнений  этих квазиклассических приближений.

Исследование А. М. Шавлукова выполнено за счет гранта Российского научного фонда, грант № 21-11-00006, https://rscf.ru/project/21-11-00006/
\medskip\\
\end{abstract}

\bigskip

\textbf{Введение.} Система квазиклассического приближения  
\begin{equation} \label{Melikhov1}
	h'_{t}+(h v)'_x=0,\quad v'_{t}+vv'_x+\alpha(h)h'_x=0
\end{equation}
в применении  к решениям нелинейного уравнения Шредингера (НУШ) с малой дисперсией 
\begin{equation} \label{Melikhov2}
	-i\varepsilon \Psi'_{t}=\varepsilon^2\Psi''_{xx}+K(|\Psi|^2)\Psi \qquad (\alpha(h)=-2K'(h), \qquad 0<\varepsilon<<1) 
\end{equation}
возникает следующим образом: подстановка в НУШ \eqref{Melikhov2} анзаца $\Psi = h ^{1/2} \exp(i\frac{\varphi}{\varepsilon})$ даёт систему двух  уравнений 
$$h '_{t}+ 2(h \varphi'_x)'_ x = 0, \quad    \varphi'_{t} + (\varphi'_x)^2 - K(h ) = \varepsilon^2\frac{( \sqrt{h })''_{xx}}{\sqrt h },$$
переход в которой к бездисперсионному пределу $\varepsilon=0$ 
\begin{equation} \label{Melikhov3}
	h '_{t}+ 2(h \varphi'_x)'_ x = 0,\quad    \varphi'_{t} + (\varphi'_x)^2 - K(h ) =0 
\end{equation}
после дифференцирования второго уравнения \eqref{Melikhov3} по переменной $x$ и замены $v=2\varphi'_x$ даёт систему \eqref{Melikhov1} c функцией  $\alpha(h)=-2K'(h).$  

\textbf{Замечание 1.} При физически осмысленных значениях $h\geqslant 0$ эта система является гиперболической при $\alpha(h)>0$ и эллиптической при $\alpha(h)<0$ c линией вырождения $h=0$. Но с математической точки зрения можно  и эллиптический случай рассматривать при $\alpha(h)>0$, считая только при этом, что  $h$ меняется в области $h\leqslant  0$.   Именно так мы в этой статье ниже и поступим, предполагая, что в малой окрестности точки $h=0$ функция $\alpha(h)$ является аналитической и представляется в ней сходящимся рядом   
\begin{equation}\label{Melikhov4}
	\alpha(h)=4+\sum_{j=1}^{\infty}\alpha_jh^j.
\end{equation}

Как в эллиптическом, так и в гиперболическом вариантах  для решений системы \eqref{Melikhov1} характерны разного рода  точки сингулярностей, вблизи которых квазиклассические приближения к решениям  \eqref{Melikhov2} теряют пригодность --- для правильного описания решений НУШ \eqref{Melikhov2} в окрестностях  этих точек сингулярностей нужно с помощью метода согласования [1] проводить отдельный анализ. Эти характерные сингулярности решений систем \eqref{Melikhov1} анализировались, в частности, в [2] --- [13].  Процитированные публикации из множества других, посвященных  таким сингулярностям, выделены существенным использованием  того факта, что преобразование годографа, при котором переменные $t$ и $x$ рассматриваются как функции $t(h,v)$, $x(h,v)$ независимых переменных  $v$ и $h$,  систему \eqref{Melikhov1} переводит в линейную систему 
\begin{equation}\label{Melikhov5}
	x'_h=vt'_h-\alpha(h)t'_v,\qquad x'_v=vt'_v-ht'_h.
\end{equation}

Часть --- см. [5] --- [13] --- из данных публикаций для описания сингулярностей квазиклассических приближений \eqref{Melikhov1} использует подход,  берущий своё начало от пионерской работы А. Х. Рахимова [14], в которой  анализировались типичные с точки зрения математической теории катастроф [15] --- [20] сингулярности решений общей $2\times2$ квазилинейной гиперболической системы  
\begin{equation}\label{Melikhov6}
	U'_t+A(U,t,x) U'_x=\Phi(U,t,x), 
\end{equation}
где $U=(u_1,u_2)$, $\Phi=(\varphi_1,\varphi_2)$ -- вектор-функции, a $A(U,t,x)$ есть $2\times2$ матрица с двумя различными вещественными собственными значениями $\lambda$, $\mu$. (В предположении о том, что зависимость этих собственных значений от инвариантов Римана $r=r(U,t,x)$, $l=l(U,t,x)$ системы (\ref{Melikhov6})  удовлетворяет неравенству $\lambda'_r\lambda'_l \mu'_r \mu'_l\neq 0$.) Суть данного подхода в применении к решениям системы \eqref{Melikhov1} состоит в следующем.      

Решения линейной системы \eqref{Melikhov5} посредством соотношений 
\begin{equation} \label{Melikhov7}
	t=B'_v, \qquad x=-B-hB'_h+vB'_v
\end{equation} 
могут быть выражены [3] через решения $B(h,v)$ скалярного уравнения 
\begin{equation}\label{Melikhov8}
	hB''_{hh}+2B'_{h}=\alpha(h)B''_{vv},
\end{equation}
которое является гиперболическим при $h>0$, и эллиптическим при  $h<0$. 
Поэтому обращение при конечных значениях $(h=h_*,v=v_*)$ и $(t=t_*,x=x_*)$ в нуль якобиана 
\begin{equation}\label{Melikhov9}
	J=x'_ht'_v-t'_hx'_v=h(t'_h)^2-\alpha(h)(t'_v)^2=h(B''_{hv})^2-\alpha(h)(B''_{vv})^2
\end{equation}
отображений $(h,v) \to (t,x)$, определяемых через локально гладкие решения уравнения \eqref{Melikhov8} формулами (\ref{Melikhov7}), сопровождается градиентной катастрофой --- обращением в бесконечность производных  $h'_x(t_*, x_*)$, $v'_x(t_*,x_*)$ и $h'_t(t_*, x_*)$, $v'_t(t_*,x_*)$ решений уравнений квазиклассического приближения \eqref{Melikhov1}.  Согласно  идеологии теории катастроф [18], для описания подобных сингулярностей общего положения можно и нужно использовать лишь разложения в ряды Тейлора гладких в точках $(h_*,v_*)$ решений $B(h,v)$ уравнения \eqref{Melikhov8} 
$$B=b_{00}+\sum\limits _{i+j=1}^{\infty}b_{ij}(h-h_*)^i(v-v_*)^j,$$
на коэффициенты $b_{ij}$ которых, помимо связей, диктуемых конкретным видом этого дифференциального уравнения,  накладывается {\it не более двух} ограничений. (По той причине, что катастрофы, отвечающие  особенностям гладкого отображения \eqref{Melikhov7}, зависят лишь от двух так называемых [18]  {\it управляющих} параметров --- в данном случае от параметров $t$ и $x$.)

Оказалось, что с точки зрения описанного выше подхода в гиперболическом случае для множества  решений \eqref{Melikhov1} типичны ровно три особенности, связанные с обращением в нуль якобиана \eqref{Melikhov9}. (Для более широкого множества решений квазилинейной гиперболической системы \eqref{Melikhov6} аналогичный вывод был сделан  А.Х. Рахимовым  ещё в [14].) В области же эллиптичности решений \eqref{Melikhov1}  типична лишь одна такая особенность, описанная в [8].  При $h_*>0$ все эти четыре особенности наследуются из типичных особенностей решений линеаризаций 
\begin{equation} \label{Melikhov10}
	h'_{t}+h_* v'_x+v_*h'_x=0,\quad v'_{t}+v_*v'_x\pm4 h'_x=0 
\end{equation}
систем \eqref{Melikhov1}:  при нулевых значениях управляющих параметров ростки отображений, описывающие  данные типичные особенности решений системы \eqref{Melikhov1}, совпадают [13] с ростками отображений, отвечающих всем типичным особенностям решений линейных уравнений $u''_{TT}\mp u''_{XX}=0$, которым эквивалентны линеаризации \eqref{Melikhov10} --- это ростки в нуле функций 
$$ A_\mu: w^{\mu+1}\quad(\mu=2,3) $$
и
$$D_4^{\pm}: w^2z \pm z^3.$$

Но в работах  [5], [12] было показано, что типичные с точки зрения теории особенностей дифференцируемых отображений градиентные катастрофы с решениями \eqref{Melikhov1}  происходят и в точках $(h_*,v_*)$ с компонентой $h_*=0$. Эти провальные сингулярности, связанные с особенностью типа сборки, из решений линейных уравнений не наследуются.  

В случае $\alpha(h)\equiv4$ уравнения \eqref{Melikhov1}   принимают вид системы 
\begin{equation} \label{Melikhov11}
	h'_{t}+(h v)'_x=0,\quad v'_{t}+vv'_x+4h'_x=0,
\end{equation}
которая в области $h\geqslant 0$ совпадает с уравнениями мелкой воды (точнее, сводится к ним простыми растяжениями). Эта система при неотрицательных (соответственно, неположительных) значениях $h$  описанным выше образом задает  квазиклассические приближения к интегрируемым методом обратной задачи рассеяния дефокусирующему НУШ  $-i\varepsilon \Psi'_{t}=\varepsilon^2\Psi''_{xx}-2|\Psi|^2\Psi$  и, соответственно, фокусирующему НУШ $-i\varepsilon \Psi'_{t}=\varepsilon^2\Psi''_{xx}+2|\Psi|^2\Psi$. Именно для системы \eqref{Melikhov11} в данной статье будет проведено строгое обоснование существования eё решений c типичными в смысле математической теории катастроф провальными асимптотиками, которые ранее на эвристическом уровне строгости для более общих систем \eqref{Melikhov1}  были описаны в работах [5], [12].   

Однако первый раздел основной части статьи начинается всё же  с описания  полных  формальных  провальных асимптотик решений  этих более  общих систем, так как приводимое в этом разделе изложение фактов и построений осуществляется методологически последовательнее и чётче, чем в [5], [12].  И к тому же при этом устраняется ряд мелких, но досадных опечаток в текстах [5], [12]. Обоснование же этих асимптотик в данной статье произведено лишь для случая  системы \eqref{Melikhov11}. Помимо фактов математической теории катастроф   это обоснование  существенным образом опирается  на частный  случай  теоремы Ю. Ф. Коробейника, сформулированной им в $\S1$ краткой публикации 1961 г. [21].

\smallskip

\textbf{1. Формальные провальные асимптотики} Для описания типичных с точки зрения теории катастроф формальных провальных асимптотик решений систем  \eqref{Melikhov1}   
будем использовать гладкие   в окрестностях точек $(h=0, v=v_*)$ решения линейного уравнения \eqref{Melikhov8}, разлагающиеся в этих точках в ряды Тейлора 
\begin{equation}\label{Melikhov12}
	B(h,v)=b_{00}+\sum\limits _{i+j=1}^{\infty}b_{ij}h^i(v-v_*)^j.
\end{equation} 

Все коэффициенты данного ряда однозначно выражаются через коэффициенты $b_{0j}$ разложения в ряд Тейлора в точке $v=v_*$ функции  $B_0(v)=B(0,v)$
\begin{equation}\label{Melikhov13}
	B_0(v)=b_{00}+\sum\limits _{j=1}^{\infty}b_{0j}(v-v_*)^j.
\end{equation}
Действительно, все коэффициенты ряда Тейлора $B=\sum_{j=0}^{\infty}h^kB_k(v)$ решения  уравнения \eqref{Melikhov8} через его главный член $B_0(v)$ рекурренто находятся по формулам

$$B_1(v)=2(B_0)''(v),\quad B_2(v)=\frac{2(B_1)''(v)}{3}+\frac{\alpha_1(B_0)''}{6},$$
$$B_{k+1}=\frac{1}{(k+1)(k+2)}(4(B_k)''_{vv}+\sum_{l=1}^k\alpha_l(B_{k-l})''(v)).$$
Из этих рекуррентных формул следуют соотношения 
$$b_{10}=4b_{02},\quad b_{11}=12b_{03},\quad b_{12}=24b_{04},\quad b_{13}=40b_{05},\quad b_{14}=60b_{05}$$ 
\begin{equation}\label{Melikhov14}
	b_{20}=32b_{04}+\frac{\alpha_1b_{02}}{3},\quad b_{21}=160 b_{05}+\alpha_1b_{03},\quad b_{22}=480b_{06}+2\alpha_1b_{04},
\end{equation}
$$b_{ij}=b_{ij}(\alpha_1, \alpha_2, \dots, \alpha_{m(i,j)}, b_{01}, b_{01}, \dots, b_{0n(i,j)}\quad(j>0).$$

Считая, что точкам $(0,v_*)$ соответствуют точки $(t_*,x_*)$, из соотношений \eqref{Melikhov7} находим, что 
\begin{equation} \label{Melikhov15}
	b_{01}=t_*,\qquad b_{00}=t_*v_*-x_*. \qquad 
\end{equation}

Обращение якобиана \eqref{Melikhov9} в нуль в точке $(0,v_*)$ для отображений \eqref{Melikhov7}, определяемых гладким в этой точке решением уравнения \eqref{Melikhov8}, означает равенство нулю коэффициента $b_{02}$ соответствующего ряда Тейлора \eqref{Melikhov12}:
\begin{equation}\label{Melikhov16}
	b_{02}=0.
\end{equation}

\textbf{Замечание 2.} На прямой $h=0$ переменные   $t$ и $x$, определяемые отображением \eqref{Melikhov7}, зависят лишь от одного параметра $v$. Поэтому при описании сингулярности общего положения, соответствующей обращению якобиана \eqref{Melikhov9} в нуль при $h=0$,   никаких других ограничений в виде равенств на коэффициенты ряда \eqref{Melikhov12} помимо тех, что следуют из результата подстановки этого ряда в уравнение \eqref{Melikhov8}, с точки зрения идеологии математической теории особенностей гладких отображений накладывать недопустимо. Именно по этой причине  мы далее считаем, что  выполнено равенство \eqref{Melikhov16} (за счет возможности вариации аргумента $v$ функций $t(0,v)$ и $x(0,v)$, являющихся управляющими параметрами соответствующей катастрофы), а коэффициент $b_{11}=12b_{03}$ в разложениях Тейлора \eqref{Melikhov12} отличен от нуля.

При таких ограничениях на эти разложения Тейлора из вида отображений \eqref{Melikhov7} и равенств \eqref{Melikhov14}, \eqref{Melikhov15} следует, что гладкие функции
\begin{equation}\label{Melikhov17}
	\tau(h,v) =t-t_*, \qquad \xi(h,v)=x-x_*-v_*(t-t_*)
\end{equation}
обладают следующими разложениями в  ряды Тейлора при $h\to0$ и $v\to v_*$:
\begin{equation}\label{Melikhov18}
	\tau =b_{11}\left[h+\frac{V^2}{4}\right]+b_{21}h^2+2 b_{12}hV+
	\sum_{i+j>2}(j+1)b_{i(j+1)}h^iV^j,
\end{equation}
\begin{equation}\label{Melikhov19}
	\xi =V\tau-3b_{20}h^2-2b_{11}hV-4b_{30}h^3-3b_{21}h^2V-2b_{12}hV^2-
	\frac{b_{11}V^3}{12}+
	\sum_{i+j>3}a_{ij}h^iV^j.
\end{equation}
Здесь
\begin{equation}\label{Melikhov20}
	V=v-v_*,
\end{equation}
а $a_{ij}$ -- постоянные, 	однозначно определяемые по коэффициентам $\alpha_k$, $b_{0k}$ разложений Тейлора \eqref{Melikhov4} и \eqref{Melikhov13}.  Так как коэффициент $b_{11}$ отличен от нуля, то из вида разложения \eqref{Melikhov18} по теореме о неявной функции следует, что в окрестности точки $(\tau=0, V=0)$ переменная $h$ определена как гладкая функция  независимых переменных $\tau$ и $V$, которая в этой точке разлагается в двойной  ряд Тейлора
\begin{equation}\label{Melikhov21}
	\begin{matrix}
		h(\tau, V)=\left(\frac{\tau}{b_{11}}+\sum\limits_{k>2}h_{0k}\tau^k \right)+V\sum\limits_{k>0}h_{1k}\tau^k+V^2\left( -\frac{1}{4}+\sum\limits_{k>0}h_{2k}\tau^k \right)+\\
		\sum\limits_{m>2,k \geqslant 0}h_{mk}V^m\tau^k
	\end{matrix}
\end{equation}
c постоянными коэффициентами $h_{mk}$, однозначно определяемыми по коэффициентам $b_{ij}$. Подстановка в разложение \eqref{Melikhov19} вместо $h$ правой части данного двойного ряда задает разложение Тейлора 
\begin{equation}\label{Melikhov22}
	\begin{matrix}
		\xi(\tau,V)=[\xi_{02}\tau^2+\sum\limits_{j>3}\xi_{0j}\tau^j] 
		+V\left[-\tau+\xi_{12} \tau^2+\sum\limits_{j>3} \xi_{1j}\tau^j\right]+\\ 
		V^2\sum\limits_{j>0} \xi_{2j} \tau^j + 
	V^3\left[\frac {5b_{11}}{12}+\sum\limits_{j>0}\xi_{3j}\tau^j\right]
	+\sum\limits_{k>3,j\geqslant 0}\xi_{kj}V^k\tau^j
\end{matrix}
\end{equation} 
гладкой в окрестности точки $(\tau=0,  V=0)$ функции $\xi$ независимых переменных $\tau$ и $V$.  Коэффициенты $\xi_{ij}$ этого ряда  также однозначно вычисляются через коэффициенты  $\alpha_k$, $b_{0k}$ рядов  \eqref{Melikhov4} и \eqref{Melikhov13}. 

Функция $\xi(0,V)$ посредством диффеоморфной в окрестности точки $W=0$ функции $V=V(W)$, которая при $W \to 0$ разлагается  в ряд Тейлора (его коэффициенты $v_j$ однозначно находятся через коэффициенты $h_{j0}$ разложения \eqref{Melikhov22})
\begin{equation}\label{Melikhov23}
	 V=V(W)=\left(\frac{12}{5b_{11}}\right)^{1/3}W+\sum_{j=2}^{\infty}v_jW^J
\end{equation}
[16, Гл. 4,\S3, Теорема 4.4],  переводится в одночлен $W^3$:
$$\xi(0,W)=\xi(0,V(W))=W^3.$$
Функцию же $\xi(\tau, V)$ этот диффеоморфизм сводит к бесконечно дифференцируемой в  окрестности точки 
$(\tau=0, W=0)$ функции $\xi(\tau, W)=\xi(\tau, V(W))$, которая, как это следует из вида ряда \eqref{Melikhov22}, при $\tau^2+W^2\to 0$ разлагается в следующий двойной ряд Тейлора:
\begin{equation}\label{Melikhov24}
	\begin{matrix}
		\xi(\tau,W)=[\xi_{02}\tau^2+\sum\limits_{j>3}\xi_{0j}\tau^j] +W\left[-\left(\frac{12}{5b_{11}}\right)^{1/3}\tau+r_{12} \tau^2+\sum\limits_{j>3} r_{1j}\tau^j\right]+\\
		W^2[\sum\limits_{j>0} r_{2j} \tau^j]+ 
		W^3[1+\sum_{j>0}r_{3j}\tau^j]
		+\sum\limits_{k>3,j>0}r_{kj}W^k\tau^j.
	\end{matrix}
\end{equation}
Постоянные коэффициенты $r_{ij}$ данного ряда также однозначно выражаются через коэффициенты $\alpha_k$ и $b_{0k}$ разложений \eqref{Melikhov4}  и \eqref{Melikhov13}. 

Равенство $\xi(0,W)=W^3$ означает, что функция  $\xi(\tau,W)$ есть деформация с параметром $\tau$ ростка одночлена $W^3$  [20, \S 13, c. 94]. Этот одночлен обладает [20, \S13, пример 13.6] также так называемой {\it R --- миниверсальной} деформацией, являющейся ростком в нуле кубического многочлена 
$W^3+\lambda_1W+\lambda_2.$
Поэтому,  в частности, существуют [20, \S 13, c.95, 100] такие гладкие отображения $U(\tau, W)$,  $\lambda_1(\tau)$, $\lambda_2(\tau),$ что $U(0,W)=W$, $\lambda_1(0)=0$, $\lambda_2(0)=0$ и
\begin{equation}\label{Melikhov25}
	\xi(\tau,W)=(U(\tau,W))^3+\lambda_1(\tau)U(\tau,W)+\lambda_2(\tau)
\end{equation}
в достаточно малой окрестности точки $(\tau=0, W=0)$. 

Непосредственной подстановкой в левую часть последнего  равенства разложения \eqref{Melikhov24}, а в его правую часть разложений Тейлора

\begin{equation}\label{Melikhov26}
	\begin{matrix}
		\lambda_k(\tau)=\sum_{j=1}\limits^{\infty}\lambda_{jk}\tau^j \qquad(k=1,2), \\
		U(\tau, W)=\sum\limits_{j>0}U_{j0} \tau^j +W\left[1+\sum\limits_{j>0}U_{j1}\tau^j\right]+
		\sum\limits_{i>1,j>0}U_{ji}W^i \tau^j
	\end{matrix}
\end{equation} 
коэффициенты $\lambda_{j1}$, $\lambda_{j2}$ и $U_{ji}$ последних рекуррентным образом однозначно находятся через постоянные коэффициенты двойного ряда Тейлора  \eqref{Melikhov24}: из равенства коэффициентов при $\tau^1$ получающихся  левой и правой частей находятся постоянные $U_{1k}$ и $\lambda_{12}=0$,  $\lambda_{11}=\left(-\frac{12}{5b_{11}}\right)^{1/3}$ --- так что при $\tau\to 0$ 
\begin{equation}\label{Melikhov27}
	\lambda_1(\tau)=\left(-\frac{12}{5b_{11}}\right)^{1/3}\tau+\sum_{j=2}^{\infty}\lambda_{j1}\tau^j, \qquad\lambda_2(\tau)=\sum_{j=2}^{\infty}\lambda_{j2}\tau^j,
\end{equation}
а из  равенства их коэффициентов при $\tau^m$ c $m>1$ --- постоянные  $\lambda_{m1}$,  $\lambda_{m2}$ и $U_{mk}$.

Согласно виду ряда \eqref{Melikhov26} при достаточно малых значениях $\lambda_2(\tau)-\xi$ и $\lambda_1(\tau)$ переменная $W$ является гладкой функцией  параметра $\tau$ и  решения $U$ кубического уравнения сборки \eqref{Melikhov25}, которая при $\tau\to 0$ и $U\to 0$ обладает разложением Тейлора  
\begin{equation}\label{Melikhov28}
	W(\tau, U)=\sum_{j>0}W_{j0} \tau^j +U\left[1+\sum_{j>0}W_{j1}\tau^j\right]+\sum_{i>1,j>0}W_{ji}U^i \tau^j,
\end{equation}
где коэффициенты $W_{ji}$  однозначно выписываются через коэффициенты $\alpha_k$ и $b_{0k}$ разложений \eqref{Melikhov4}, \eqref{Melikhov12}. 

Подводя итоги вышесказанному в этом разделе статьи, констатируем, что   доказана 

\textbf{Теорема 1.}

Допустим, что  $\alpha(h)$ представляет собой локально аналитическую функцию, представляемую сходящимся рядом \eqref{Melikhov4}. И пусть $B(h,v)$ есть гладкое в достаточно малой окрестности точки $(h=0,  v=v_*)$ решение линейного  дифференциального уравнения \eqref{Melikhov8}, у которого в разложении Тейлора \eqref{Melikhov12} $b_{02}=0$, $b_{11}=12b_{03}\neq0$, а $t_*$ и $x_*$ --- значения гладких отображений \eqref{Melikhov7}  при $h=0$  и $v=v_*$.  Тогда данные отображения \eqref{Melikhov7} задают сингулярные в точке $(t_*,x_*)$ решения системы \eqref{Melikhov1}.  После перехода к независмым переменным 
\eqref{Melikhov17},   сдвига \eqref{Melikhov20} и локального диффеоморфизма $V=f(W)$  c разложением Тейлора  \eqref{Melikhov23} компоненты $v$ этих сингулярных решений при $t\to t_*$ и $x\to x_*$ задаются любым из трёх корней $U(\tau,\xi)$ кубического уравнения сборки  \eqref{Melikhov25} посредством двух гладких функций $\lambda_1(\tau)$ и $\lambda_2(\tau)$ c разложениями Тейлора  \eqref{Melikhov27} и гладкой в точке $(\tau=0,U=0)$ функции $W=W(\tau,U)$, разлагающейся в двойной ряд Тейлора \eqref{Melikhov28}.  Компоненты  $h$ данных сингулярных решений системы \eqref{Melikhov1} при этом определяются первым из отображений \eqref{Melikhov7} как гладкие в точке $(\tau=0,V=0)$ функции, которые разлагаются в двойной ряд Тейлора \eqref{Melikhov21}.

Таким образом, асимптотики решений \eqref{Melikhov1}, соответствующие типичной провальной особенности сборки гладких отображений   \eqref{Melikhov7}, описаны со сверхстепенной точностью.

\textbf{Замечание 3.} Решение $U$  канонического уравнения катастрофы сборки \eqref{Melikhov25}  для всех значений $\lambda_2-\xi$ единственно и гладко по обоим управляющим параметрам $\lambda_1$,  $\lambda_2-\xi$ лишь при положительных $\lambda_1$.  Однако для неположительных значений $\lambda_1$  однозначность и гладкость имеет место лишь вне промежутка 
\begin{equation}\label{Melikhov29}
	|\lambda_2-\xi|<\left(\frac{-4(\lambda_1)^3}{27}\right)^{1/2}.
\end{equation}
На $\xi,\tau$-плоскости границы этого интервала представляют собой две кривые $\xi_{\pm,c}(\tau)$, которые изначально исходят из точки $(\xi=0,\tau=0)$  в полуплоскость $\tau b_{11}<0$ и при $\tau \to 0$ в главном порядке задаются формулами 
\begin{equation}\label{Melikhov30}
	\xi_{\pm,c}(\tau)=\pm\frac{4}{3}\left|\frac{\tau^{3}}{5b_{11}}\right|^{1/2}(1+o(1)).
\end{equation} 
Во всех внутренних точках промежутка \eqref{Melikhov29} кубическое уравнение  \eqref{Melikhov25} имеет три различных между собой корня. Как следует из разложений \eqref{Melikhov21} и \eqref{Melikhov22}  две кривые $\xi_{\pm,z}(\tau)$ обращения в нуль компоненты $h$ сингулярных решений системы  \eqref{Melikhov1}, описываемых в теореме 1,  при малых $\tau$ также  расположены в полуплоскости $\tau b_{11}<0$ и при $\tau\to 0$ обладают асимптотиками     
\begin{equation}\label{Melikhov31}
	\xi_{\pm,z}(\tau)=\pm\frac{4}{3}\left|\frac{\tau^{3}}{b_{11}}\right|^{1/2}(1+o(1)).
\end{equation}
Из сравнения асимптотик \eqref{Melikhov30}  и \eqref{Melikhov31} следует, что, по крайней мере, при малых $\tau$  данные кривые провала $\xi_{\pm,z}(\tau)$ лежат в области единственности решения уравнения сборки \eqref{Melikhov25}.  Надо иметь в виду, что по разные стороны от этих кривых провала решения системы \eqref{Melikhov1}, описанные в теореме 1,  в применении к задаче описания квазиклассических приближений  к решениям НУШ \eqref{Melikhov2} могут иметь смысл лишь для одного из двух вариантов  --- эллиптического или гиперболического --- данной системы.  Кстати, решения уравнения движения одномерного изоэнтропического газа с произвольным давлением, совпадающие с гиперболическим вариантом системы \eqref{Melikhov1}, частным случаем которых являются уравнения мелкой воды \eqref{Melikhov11}, физически осмыслены  лишь при неотрицательных значениях плотности (для уравнений мелкой воды \eqref{Melikhov11}  --- отклонения уровня жидкости от дна) $h$. Но во всяком случае методическим достоинством данной статьи её авторы как раз считают <<абстрагирование от физического смысла>>  при обосновании формальных асимптотик решений системы \eqref{Melikhov11} сразу же для эллиптического и гиперболического вариантов.  Первоначальные же наши попытки проведения такого обоснования отдельно для каждой из областей $h\geqslant 0$ и $h\leqslant 0$ встретились со  значительными  затруднениями.

Доказанная {\bf теорема 1} носит несколько условный характер, поскольку априори неочевидна даже непустота   множества гладких  в точке $(h=0,v=v_*)$ решений  уравнения \eqref{Melikhov8}, имеющих разложения Тейлора \eqref{Melikhov12} с коэффициентами $b_{02}=0$, $b_{11}=12b_{03}\neq0$. Но для частного случая $\alpha(h)\equiv4$ можно доказать, что этому множеству принадлежат все локально аналитические в точке $(h=0,v=v_*)$ решения $B(h,v)$, определяемые любыми  функциями $B(0,v)$ со сходящимися при достаточно малых значениях $V$ рядами Тейлора 
$$B(0,v)=v_*t_*-x_* +t_*V+ b_{03}V^3+\sum_{j=4}^{\infty}b_{0j}V^j, \qquad b_{03}\neq 0.$$  
Это  возможно из-за сводимости заменой $B(h,v)=\frac{C(h,v)}{h}$ уравнения \eqref{Melikhov8} к уравнению 
\begin{equation}\label{Melikhov32}
	h C''_{hh}=\alpha(h)C''_{vv},
\end{equation}
для которого согласно теореме, сформулированной   Ю. Ф. Коробейником в кратком сообщении [21], при $\alpha(h)\equiv4$ из аналитичности в точке $u=u_*=\frac{v_*}{2}$ функции $g_1(u)$, однозначно определяющей формальное решение \eqref{Melikhov32} в виде ряда  
\begin{equation}\label{Melikhov33}
	C(h,v)=G(h,u)=g_1(u)h+\sum\limits_{k=2}^\infty g_k(u)h^k\qquad(u=\frac{v}{2}),
\end{equation}
следует, что данный ряд  есть аналитическая функция переменных $h$ и $u$ в некотором бикруге 
\begin{equation} \label{Melikhov34} 
	D(R_1, R):=\{h,u\in\mathbb C, \,\, |h|<R_1, \, |u-u_*|<R\}.
	\end{equation}

\textbf{Замечание 4.} Упомянутая теорема  в [21] касается решений более широкого класса уравнений вида $\frac{\partial^k G}{\partial u^k}=h^m\frac{\partial^n G}{\partial h^n}$  с $n>m$.  К сожалению, нам нигде не удалось найти опубликованное доказательство этой теоремы Ю. Ф. Коробейника. Но для случая $k=n=2$ и $m=1$, то есть для решений уравнения
\begin{equation}\label{Melikhov35}
	hG''_{hh}=G''_{uu},  
\end{equation}
её доказательство нами  было получено независимо.  Следующий, завершающий раздел статьи посвящён изложению данного доказательства.

\textbf{2. Теорема Коробейника.} 

Отметим, что коэффициенты $g_{k+1}(u)$ при $k\ge 1$ формального решения в виде ряда \eqref{Melikhov33}  уравнения \eqref{Melikhov35} выражаются через производные $g_1^{(2k)}(u)$ порядка $2k$ коэффициента $g_1(u)$ 
в некоторой окрестности точки $u=u_*$ равенствами
$g_{k+1}(u)=\frac{g_1^{(2k)}(u)}{k!(k+1)!}.$
Поэтому в дальнейшем пойдет речь о формальном решении уравнения \eqref{Melikhov35} в виде ряда
\begin{equation} \label{Melikhov36}
	G(h,u)=g_1(u)h+\sum\limits_{k=1}^\infty\frac{1}{k!(k+1)!}g_1^{(2k)}(u)h^{k+1}.
\end{equation}

Приведём точную формулировку интересующей нас  части теоремы Ю. Ф. Коробейника. 

\textbf{Теорема 2 (Ю. Ф. Коробейник, 1961 г.).} Пусть $R, R_1$ --- произвольные вещественные положительные числа, $u_*$ --- произвольное комплексное число, $G(h,u)$ ---  формальное решение
в виде ряда \eqref{Melikhov36} уравнения \eqref{Melikhov35},
где функция $g_1(u)=g_1(Re\, u+iIm\, u)$  бесконечно дифференцируема в круге $\{u\in\mathbb C,\,\, |u-u_*|<R\}$ как функция двух вещественных переменных $Re\, u$, $Im\, u$.
Для того, чтобы функция $G(h,u)$ была аналитической в бикруге \eqref{Melikhov34} 
и ряд \eqref{Melikhov36} сходился по $h$ в круге $\{h\in\mathbb C, \,\,|h|<R_1\}$ при любом $u$ из  круга 
$\{u\in \mathbb C, \,\,|u-u_*|<R\}$, необходимо и достаточно, чтобы функция $g_1(u)$  была аналитической в круге $\{u \in \mathbb{C},  |u-u_*|<R+2\sqrt{R_1}\}$.

Переходим теперь к её доказательству. Вначале докажем следующее вспомогательное утверждение.

\textbf{Лемма.} Пусть $f(z)$ --- функция, аналитическая в круге $\{z\in\mathbb C,\,\,|z|<r\}$, $r\in (0,\infty)$. Для произвольных $r_0\in [0,r)$ и  $\varepsilon\in(0,r-r_0)$ 
существует такая  положительная постоянная $C(\varepsilon)$, что при всех натуральных $n$ в  круге $|z|\leqslant  r_0$ справедлива  оценка
$$	|f^{(n)}(z)|\leqslant  C(\varepsilon)\frac{n!(r-\varepsilon)}{(r-r_0-\varepsilon)^{n+1}}.$$

\textbf{Доказательство леммы.} Вследствие интегральной формулы Коши 
для любого $z\in\mathbb C$, лежащего в круге $|z|\le r_0$, при всех натуральных $n$ выполняется равенство
$$ f^{(n)}(z)=\frac{n!}{2\pi i}\int\limits_{|t|=r-\varepsilon}\frac{f(t)}{(t-z)^{n+1}}dt. $$
Отсюда следует, что
$$ |f^{(n)}(z)|\leqslant\frac{n!}{2\pi} 2\pi(r-\varepsilon)\max\limits_{|t|=r-\varepsilon}\frac{|f(t)|}{|t-z|^{n+1}}\leqslant C(\varepsilon)\frac{n!(r-\varepsilon)}{(r-r_0-\varepsilon)^{n+1}}, $$
где $C(\varepsilon)=\max\limits_{|t|=r-\varepsilon}|f(t)|$. Лемма доказана.

Докажем достаточность в теореме 2. Предположим, что функция $g_1(u)$  аналитична в круге $\{u \in \mathbb{C},  |u-u_*|<R+2\sqrt{R_1}\}$. 
Покажем, что для любого  $\delta \in(0,R)$ ряд  в правой части \eqref{Melikhov36} сходится равномерно на бикруге
\begin{equation} \label{Melikhov37}
	E(\delta)=\{ h, u \in \mathbb{C}, \quad |h|\leqslant R_1, \quad |u-u_*|\leqslant R-\delta\}.
\end{equation}

Рассмотрим величины
$$\beta_k=\sup\left\{\frac{|h|^{k+1}}{k!(k+1)!}|g_1^{(2k)}(u)|,\,\, |h|\leqslant R_1, |u-u_*|\leq R-\delta\right\}.$$
В силу только что доказанной леммы, в которой надо положить $f(z)=g_1(z+u_*)$, $z=u-u_*$, $r=R+2\sqrt{R_1}$, $r_0=R-\delta$, для любого $\varepsilon \in (0, \delta)$ существует такая положительная постоянная $C(\varepsilon)$, что при всех натуральных $k$ на круге $|u-u_*|\leqslant R-\delta$ имеет место оценка
$$ |g_1^{(2k)}(u)|\leqslant \frac{C(\varepsilon) (2k)! (R+2\sqrt{R_1}-\varepsilon)}{(2\sqrt{R_1}+\delta-\varepsilon)^{2k+1}}. $$
Значит,
$$	\beta_k\leqslant (R_1)^{k+1}\frac{C(\varepsilon) (2k)! (R+2\sqrt{R_1}-\varepsilon)}{k!(k+1)!(2\sqrt{R_1}+\delta-\varepsilon)^{2k+1}}=:\alpha_k.$$
Ряд $\sum\limits_{k=1}^\infty\alpha_k$ сходится. Действительно, для любого $k\geqslant 1$
$$ \frac{\alpha_{k+1}}{\alpha_k}=\frac{R_1(2k+1)(2k+2)}{(k+1)(k+2)(2\sqrt{R_1}+\delta-\varepsilon)^2} $$
и 
$$ \lim\limits_{k\to\infty}\frac{\alpha_{k+1}}{\alpha_k}=\frac{R_1}{\left(\sqrt{R_1}+\frac{\delta-\varepsilon}{2}\right)^2}<1. $$
Таким образом, ряд \eqref{Melikhov36} равномерно сходится на множестве $E(\delta)$ (см.\eqref{Melikhov37}) для любого $\delta\in(0,R)$. Поэтому в бикруге \eqref{Melikhov34} функция $G(h,u)$ действительно аналитична.

Перейдем теперь к доказательству необходимости. Предположим, что функция $G(h,u)$ аналитична в бикруге \eqref{Melikhov34} и
ряд \eqref{Melikhov36} сходится по $h$ в круге $\{h\in\mathbb C, \,\,|h|<R_1\}$ для любого $u$, удовлетворяющего неравенству $|u-u_*|<R$.
Поскольку $g_1(u)=\frac{\partial G}{\partial h}(0,u)$, $|u-u_*|<R$, то функция $g_1(u)$ аналитична в круге $\{u\in\mathbb C, \,|u-u_*|<R\}$. Покажем, что её можно аналитически продолжить в круг $\{u\in\mathbb C,\,|u-u_*|<R+2\sqrt{ R_1}\}$ c помощью степенного ряда
\begin{equation} \label{Melikhov38}
	g_1(u)+\sum\limits_{j=1}^\infty\frac{g_1^{(j)}(u)}{j!}h^j.
\end{equation}
Прежде всего, вследствие представления \eqref{Melikhov36} для любого $u\in\mathbb C$ такого, что $|u-u_*|<R$, при каждом натуральном $k$ справедливо равенство
$$
g_1^{(2k)}(u)=k!\frac{{\partial}^{k+1}G}{\partial h^{k+1}}(0,u).
$$ 
Зафиксируем числа $t\in(0,R)$, $t_1\in (t,R)$ и числа $r_0\in(0,2\sqrt{R_1})$, $\rho\in(r_0,2\sqrt{R_1})$. 
Из равенств
$$ \frac{{\partial}^{k+1}G}{\partial h^{k+1}}(0,u)=\frac{(k+1)!}{2\pi i}\int\limits_{|h|=\rho^2/4}\frac{G(h,u)}{h^{k+2}}dh $$
следует, что для любого $u\in\mathbb C$, для которого $|u-u_*|\leqslant t_1$, при всяком натуральном  $k\geqslant 1$
\begin{equation} \label{Melikhov39}
	|g_1^{(2k)}(u)|\leqslant \frac{k! (k+1)!M(\rho, t_1)}{(\rho^2/4)^{k+1}}=\frac{4^{k+1} k!(k+1)!}{\rho^{2k+2}}M(\rho, t_1),
\end{equation}
где
$$ M(\rho,t_1):=\max\{|G(h,u)|,\,\, |h|\leqslant \rho^2/4,\, |u-u_*|\leqslant t_1\}. $$
Функция $g^{(2k+1)}(u)$  является производной $g^{(2k)}(u)$. Поэтому, используя формулу Коши
$$ g^{(2k+1)}(u)=\frac{1}{2\pi i}\int\limits_{|v-u_*|=t_1}\frac{g^{(2k)}(v)}{(v-u)^2}dv, $$
получим, что для любого натурального $k\geqslant 1$ и всякого  $u\in\mathbb C$ такого, что $|u-u_*|\leqslant t$, выполняется неравенство
\begin{equation} \label{Melikhov40}
|g_1^{(2k+1)}(u)|\leqslant\frac{t_14^{k+1}k!(k+1)!M(\rho, t_1)}{(t_1-t)^2\rho^{2k+2}}.
\end{equation}
Вследствие неравенства \eqref{Melikhov39} для любого натурального $k\geqslant 1$
$$ \sup\left\{\frac{|g^{(2k)}(u)|}{(2k)!}|h|^{2k},\,\,|u-u_*|\leqslant t_1,\, |h|\le r_0\right\}\le M(\rho, t_1)\frac{4^{k+1} k!(k+1)!}{(2k)!\rho^{2k+2}}(r_0)^{2k}. $$
Отсюда следует (см. доказательство достаточности), что ряд $\sum\limits_{k=1}^\infty\frac{g^{(2k)}(u)}{(2k)!}h^{2k}$ для любого $t_1\in(0,R)$ и каждого $u\in\mathbb C$ из круга $|u-u_*|\leqslant t_1$, сходится равномерно на круге $\{h\in\mathbb C,\,\,|h|\le r_0\}$. Значит, он сходится для любых $u, h\in\mathbb C$ таких, что $|u-u_*|<R$, $|h|<2\sqrt{R_1}$. Аналогично, с помощью неравенств \eqref{Melikhov40}, показывается, что ряд $\sum\limits_{k=1}^\infty\frac{g^{(2k+1)}(u)}{(2k+1)!}h^{2k+1}$ также сходится для всех $u, h\in\mathbb C$, таких, что  $|u-u_*|<R$, $|h|<2\sqrt{R_1}$. Таким образом, для всех $u, h\in\mathbb C$, удовлетворяющих неравенствам $|u-u_*|<R$, $|h|<2\sqrt{R_1}$, ряд \eqref{Melikhov38} сходится.
При этом его сумма  для каждого $u\in\mathbb C$ из круга $|u-u_*|<R$ аналитически продолжает функцию $g_1(w)$ в круг $\{w\in\mathbb C,\,\, |w-u|<2\sqrt{R_1}\}$. Поэтому $g_1(u)$ аналитически продолжима в больший круг $\{u\in\mathbb C,\,\,|u-u_*|< R+2\sqrt{R_1}\}$.

\textbf{Теорема 2} доказана.

\medskip

Если зафиксировать сумму $R+2\sqrt{R_1}$, а положительные числа $R$ и $R_1$  менять, то можно получить более широкую область, в каждой точке  которой ряд Тейлора  $G(h,u)$ вида \eqref{Melikhov36} сходится (к $G(h,u)$) и функция $G(h,u)$ аналитическая.  Эта область является объединением бикругов \eqref{Melikhov34} с центром в точке $h=0, u=u_*$,  о которых идет речь в предыдущей теореме.

Зафиксируем положительное число $R_0$. Определим область в $\mathbb C^2$
$$ P(R_0)=\{ h,u\in\mathbb C, \,\,|u-u_*|+2\sqrt{|h|}< R_0\}.$$
Ясно, что $$P(R_0)=\bigcup\limits_{R, R_1>0,\,\, R+2\sqrt{R_1}=R_0} D(R_1, R).$$ 
Предположим, что  функция $g_1(u)$ аналитична в круге $\{u\in\mathbb C,\,|u-u_*|<R_0\}$. Пусть функция $G(h,u)$ задаётся рядом \eqref{Melikhov36}. Возьмем точку $(h,u)$ в $P(R_0)$. Найдутся числа $R, R_1>0$, для которых $R+2\sqrt{R_1}=R_0$ и $(h,u)$ принадлежит бикругу $D(R_1,R)$. По {\bf теореме 2} ряд в правой части формулы  \eqref{Melikhov36} сходится в каждой точке $D(R_1,R)$, функция $G(h,u)$ аналитична в $D(R_1,R)$ и является решением уравнения \eqref{Melikhov35}. Получили такое 

\textbf{\bf Следствие.} Пусть функция $g_1(u)$ аналитична в круге $\{u\in\mathbb C,\,|u-u_*|<R_0\}$, где $R_0$ --- положительное число. Тогда ряд в правой части \eqref{Melikhov36} сходится в каждой точке из $P(R_0)$ к функции $G(h,u)$, являющейся решением уравнения \eqref{Melikhov35}, аналитическим в области $P(R_0)$.

\textbf{Замечание 5.} А. И. Янушаускасом в [22], [23, Гл. 6, \S5] описывались аналитические решения ряда эллиптических и вырождающихся уравнений в частных производных. Эти решения для части из рассмотренных в [22], [23] уравнений строятся в виде рядов по одной из независимых переменных, сходимость которых обосновывается с помощью метода мажорант. В частности, в этих публикациях изучена структура решений уравнения 
\begin{equation}\label{Melikhov41}
	s(v_k)''_{ss}+(k+1)(v_k)'_s+(v_k)''_{zz}=0,  
\end{equation}
(в  нумерации из [22] уравнение -- (16) для случая оператора $L=\frac{\partial^2} {{\partial}
z^2}$), при $k=1$ сводящегося к исследуемому нами уравнению \eqref{Melikhov35}. Но для уравнения \eqref{Melikhov41} в [22] нет обоснования сходимости соответствующих рядов, не описана область, в которой рассматриваемые ряды сходятся, не указана её зависимость  от  определяющей функции $v_0^{(k)}(z)$. Доказанные нами {\bf теорема 2} и её {\bf следствие} позволяют проследить связь областей специального вида, в которых решение $G(h,u)$ уравнения \eqref{Melikhov35} и коэффициент $g_1(u)$ аналитичны. Непосредственно же из формулировок [22], [23, Гл.4, \S5] не видно, например, как извлечь обоснованность утверждения, сделанного выше в абзаце над {\bf замечанием 4}.

\smallskip

\textbf{Заключение.} Правдоподобным выглядит предположение о том, что в случае локальной аналитичности в точке $u=2v_*$ функции $g_1(u)$ формальные асимптотические решения  вида \eqref{Melikhov33}  уравнения \eqref{Melikhov32} c произвольным коэффициентом $\alpha(h)$, представимым сходящимся рядом \eqref{Melikhov4}, в некотором бикруге  \eqref{Melikhov34} по-прежнему будет аналитическими функциями переменных $h$ и $u$. 

Строгое доказательство справедливости этой гипотезы устранило бы условный характер {\bf теоремы 1} --- см. начало абзаца над   {\bf замечанием 4}  --- и для случая произвольной аналитической в окрестности точки $h=0$ функции $\alpha(h)$, принимающей в нуле значение 4.

Авторы заявляют, что стороны не имеют конфликта интересов. 

\bigskip

\begin{center}{СПИСОК ЛИТЕРАТУРЫ}\end{center}{\small
\lit{1}{Ильин А.М.}{Согласование асимптотических разложений решений краевых задач // М. Наука. 1989. С. 1 -- 336.}
\lit{2}{Гуревич А.В., Шварцбург А.Б.}{Нелинейная теория распространения радиоволн в ионосфере // М.: Наука. 1973. С. 1 -- 272.}
\lit{3}{Шварцбург А.Б.}{Геометрическая оптика в нелинейной теории волн // М. Наука. 1976. С. 1 -- 119.}
\lit{4}{Жданов C.Л., Трубников Б.А.}{Квазигазовые неустойчивые среды // М. Наука. 1991. С. 1 -- 174.}
\lit{5}{Кудашев В.Р., Сулейманов Б.И.}{Особенности некоторых типичных процессов самопроизвольного падения интенсивности в неустойчивых средах // Письма в ЖЭТФ. 1995. Т. 65. №4. С. 358 -- 363.}
\lit{6}{Кудашев В.Р., Сулейманов Б.И.}{Влияние малой диссипации на процессы зарождения одномерных ударных волн // ПММ. 2001. Т. 3. №3. С. 456 -- 466.}
\lit{7}{Гарифуллин Р.Н., Сулейманов Б.И.}{От слабых разрывов к бездиссипативным ударным  волнам // Журн. Экс. и Теор. Физ. 2010. Т. 137. №1. С. 149 -- 164.}
\lit{8}{Dubrovin B., Grava T., Klein C.}{On universality of critical behaviour in the critical behaviour in the focusing nonlinear Schr\"odinger equation, elliptic umbilic catstrophe and the tritonque to the Painlev\'e-I equation // J. Nonlinear Sc. 2009. V. 19. № 1. P. 57 -- 94.}
\lit{9}{Konopelchenko B.G., Ortenzi G.}{Quasi-classical approximation in vortex filament dynamics. Integrable systems, gradient catastrophe, and flutter // Stud. Appl. Math. 2013. V. 130. № 2. P. 167 -- 199.}
\lit{10}{Konopelchenko B.G., Ortenzi G.}{Jordan form, parabolicity and other features of change of type transition for hydrodynamic type systems // J. Phys. A. 2017. V. 50. № 21. Art. No. 215205.}
\lit{11}{Богаевский И.А., Туницкий Д.В.}{Особенности многозначных решений квазилинейных гиперболических систем // Тр. МИАН. 2020. Т. 308. С. 76 -- 87.}
\lit{12}{Сулейманов Б.И., Шавлуков A.М.}{Типичная провальная особенность сборки  решений уравнений движения одномерного изоэнтропического газа // Изв. РАН. Серия физ. 2020. Т.84. №5. С. 664 -- 666.}
\lit{13}{Сулейманов Б.И., Шавлуков A.М.}{О наследовании решениями уравнений движения изоэнтропического газа типичных особенностей решений линейного волнового уравнения // Матем. заметки. 2022. Т. 112. № 4. С. 625 -- 640.}
\lit{14}{Рахимов А.Х.}{Особенности римановых инвариантов // Функц. анализ и его прил. 1993. Т. 27. № 1. С. 46 -- 59.}
\lit{15}{Брёкер Т., Ландер Л.}{Дифференцируемые ростки и катастрофы // М. Мир. 1977. С. 1 -- 208.}
\lit{16}{Постон Т., Стюарт И.}{Теория катастроф и ее приложения // М. Мир. 1980. С. 1 -- 617.}
\lit{17}{Арнольд В.И., Варченко А.Н., Гусейн-Заде С.М.}{Особенности дифференцируемых отображений. Том 1. Классификация критических точек, каустик и волновых фронтов // М. Наука. 1982. С. 1 -- 304.}
\lit{18}{Гилмор Р.}{Прикладная теория катастроф. Книга 1 // М. Мир. 1984. С. 1 -- 349.}
\lit{19}{Алексеев Ю.К., Сухоруков В.П.}{Введение в теорию катастроф // М. МГУ. 2000. С. 1 -- 182.}
\lit{20}{Седых В.Д.}{Математические методы теории катастроф // М. МЦНМО. 2021. С. 1 -- 224.}
\lit{21}{Коробейник Ю.Ф.}{Об аналитических решениях одного класса уравнений в частных производных // Докл. АН СССР. 1961. Т. 140. №6. С. 1248 -- 1251.}
\lit{22}{Янушаускас А.И.}{Структурные свойства решений некоторых аналитических уравнений с частными производными // Дифференц. уравнения. 1981.  Т. 17. №.1. С. 182 -– 194.}
\lit{23}{Янушаускас А.И.}{Аналитические и гармонические функции многих переменных // Новосибирск. Наука. 1981. С. 1 -- 183.}

\pagebreak

} 
\newpage

СВЕДЕНИЯ ОБ АВТОРАХ (копия карточки автора для каждого автора в отдельности)\\

1.	ФИО: Мелихов Сергей Николаевич (Melikhov Sergej Nikolaevich)

2.	Дата рождения: 05.03.1960

3.	Место работы: Южный федеральный университет (Southern Federal University); Южный математический институт ВНЦ РАН (Southern Mathematical Institute of VSC of RAS)

4.	Занимаемая должность: профессор кафедры алгебры и дискретной математики; ведущий научный сотрудник

5.	Ученое звание и степень: доктор физико-математических наук

6.	Служебный адрес: Ростов-на-Дону, ул. Б. Садовая, 105/42; Владикавказ, ул. Маркуса, 22

7.	Телефон (с кодом города): +7 863 297 51 11; +7 867 223 0051

8.	Домашний адрес: Ростов-на-Дону, пер. Чаленко, 17/1, кв. 158

9.	Телефон (с кодом города): +7 918 895 86 02

10.	Электронный адрес: snmelihov@yandex.ru

11.	Факс: +7 863 297 51 13

12.	Основные направления научных исследований: ряды экспонент, операторы свертки, пространства аналитических функций и операторы в них, линейные непрерывные правые обратные операторы, проективные описания, ультрараспределения, оператор обратного сдвига, циклические векторы, инвариантные подпространства. 

13. Название статьи: Типичные провальные асимптотики квазиклассических приближений к решениям нелинейного уравнения Шрёдингера. 
(Typical dropping asymptotics of quasiclassical approximations to solutions of the nonlinear Schr\"odinger equation.)

14. УДК: 517.956

15. Раздел (рубрика), к которому относится статья:\\ Уравнения с частными производными. 

\medskip

1.	ФИО: Сулейманов Булат Ирекович (Suleimanov Bulat Irekovich)

2.	Дата рождения: 27.05.1958

3.	Место работы: Институт математики с вычислительным центром (Institute of Mathematics with Computing Centre)

4.	Занимаемая должность: ведущий научный сотрудник

5.	Ученое звание и степень: доктор физико-математических наук

6.	Служебный адрес: Уфа, ул. Чернышевского, 112

7.	Телефон (с кодом города): +7 347 272 59 36

8.	Домашний адрес: Уфа, Проспект Октября, 12, кв.37

9.	Телефон (с кодом города): +7 937 318 41 52

10.	Электронный адрес: bisul@mail.ru

11.	Факс: +7 347 272 59 36

12.	Основные направления научных исследований: интегрируемые нелинейные уравнения, асимптотики решений нелинейных уравнений, теория катастроф. 

13. Название статьи: Типичные провальные асимптотики квазиклассических приближений к решениям нелинейного уравнения Шрёдингера. 
(Typical dropping asymptotics of quasiclassical approximations to solutions of the nonlinear Schr\"odinger equation.)

14. УДК: 517.956

15. Раздел (рубрика), к которому относится статья:\\ Уравнения с частными производными. 

\medskip

1.	ФИО: Шавлуков Азамат Мавлетович (Shavlukov Azamat Mavletovich)

2.	Дата рождения: 15.09.1994

3.	Место работы: Институт математики с вычислительным центром (Institute of Mathematics with Computing Centre)

4.	Занимаемая должность: инженер-исследователь

5.	Ученое звание и степень: -

6.	Служебный адрес: Уфа, ул. Чернышевского, 112

7.	Телефон (с кодом города): +7 347 272 59 36

8.	Домашний адрес: Уфа, ул. Кирова, 43/1, кв.67

9.	Телефон (с кодом города): +7 987 601 61 80

10.	Электронный адрес: aza3727@yandex.ru

11.	Факс: +7 347 272 59 36

12.	Основные направления научных исследований: теория особенностей дифференцируемых отображений. 

13. Название статьи: Типичные провальные асимптотики квазиклассических приближений к решениям нелинейного уравнения Шрёдингера. 
(Typical dropping asymptotics of quasiclassical approximations to solutions of the nonlinear Schr\"odinger equation.)

14. УДК: 517.956

15. Раздел (рубрика), к которому относится статья:\\ Уравнения с частными производными.

Название организации указывать без сокращений, телефоны и электронный адрес указывать обязательно.

\end{document}